\documentclass{camera}
\usepackage{graphicx}
\usepackage{amssymb}
\begin{document}

%
\title{CLUSTERS OF GALAXIES: A FUNDAMENTAL PILLAR OF COSMOLOGY.}

%
\author{Africa Castillo-Morales$^{1}$ \And Sabine Schindler$^{2}$}

%
\organization{
             $^{1}$Dpto. F\'{\i}sica Te\'orica y del Cosmos,
             Universidad de Granada,
             Avda. Fuentenueva s/n, 18002 Granada, Spain

             $^{2}$Institut f\"ur Astrophysik,
             Leopold-Franzens-Universit\"at Innsbruck,
             Technikerstr. 25, A-6020 Innsbruck, Austria}
\maketitle

%
\begin{abstract}
Clusters of galaxies are used in a variety of ways to do cosmology. Some of them are presented here. Their X-ray emitting gas allows us to determine the baryon fraction, dark matter distribution and the matter density $\Omega_{m}$ of the universe. Another interesting component is relativistic particles whose radio emission provide the measure of the magnetic fields ($\approx \mu G$) in the intra-cluster medium (ICM). The observation of distant clusters of galaxies is also important for cosmology. In particular the new X-ray satellites CHANDRA and XMM yield exciting new results for galaxy cluster physics and cosmology.

\end{abstract}

\section{Introduction}
Clusters of galaxies have been shown to be excellent tools for cosmological research. They provide a variety of different ways to determine cosmological parameters like the mean density $\Omega_{m}$ of the universe, the baryon density $\Omega_{baryon}$, the cosmological constant $\Lambda$ or the Hubble constant $H_{0}$. Clusters of galaxies are important for several reason: (1) They are the largest bound systems in the universe. With sizes of a few Mpc and masses around $10^{15}M_{\odot}$ they constitute a considerable fraction of the universe. (2) They are closed systems. Moreover, the crossing time - the time it takes a galaxy to move from one end of the cluster to the other - is of the order of a Hubble time. As no matter can leave the deep cluster potential well, all the metals that have been processed inside the cluster must still be present; {\it{i.e}} all the traces of formation processes in clusters are still observable which provide insights into the early universe. (3) Clusters are observable out to large redshifts. Comparison of properties of distant and nearby clusters yield information about evolutionary effects, which are predicted to be different in different cosmological scenarios. (4) They can be used to trace the mass distribution in the universe, which is another quantity to constrain cosmological models.
Throughout this article we use $H_{0}=50$ km/s/Mpc.

\section{Cluster Components}

Clusters of galaxies are formed by different observable components: hundreds of galaxies, hot gas between the galaxies and sometimes relativistic particles. These components are emitting in different wavelengths from radio to X-rays. With the combination of all these observations at different frequencies and also with theoretical models we can derive a good understanding of these massive objects.
Clusters were first discovered as associations of hundreds to thousands of galaxies. These galaxies are not at rest in the potential well, but move around with velocities of the order of 1000 km/s. The galactic mix of morphological types of galaxies in clusters differ from the mix in the field (Dressler 1980~\cite{Dre}): in clusters an excess of elliptical galaxies is visible, which is an indication of interaction. This interaction can be between galaxies or between galaxies and another component: the intra-cluster gas.

This gas fills all the space between the galaxies. The density of the gas is relatively low with $10^{-4}-10^{-2}$ particles/$cm^{3}$, but the temperature is high at $10^{7}-10^{8}$K. Such high temperatures are in good agreement with the depth of the potential wells of clusters. Gas with these specifications emits thermal bremsstrahlung in the X-ray range, which makes it a particularly interesting cluster component because we can observe it with X-ray satellites. These observations provide spatial information about the morphology of clusters as well as spectral information which can be used to determine gas temperatures. X-ray spectra also show lines which correspond to metallicities of about one third of the solar value (Fukazawa {\it{et al.}} 1998~\cite{Fuk}). This is an indication that the gas cannot be only of primordial origin, but at least part of it must have been processed in cluster galaxies and expelled later from the galaxy potential well to the ICM.
\begin{figure}
\centering
\includegraphics[width=6cm]{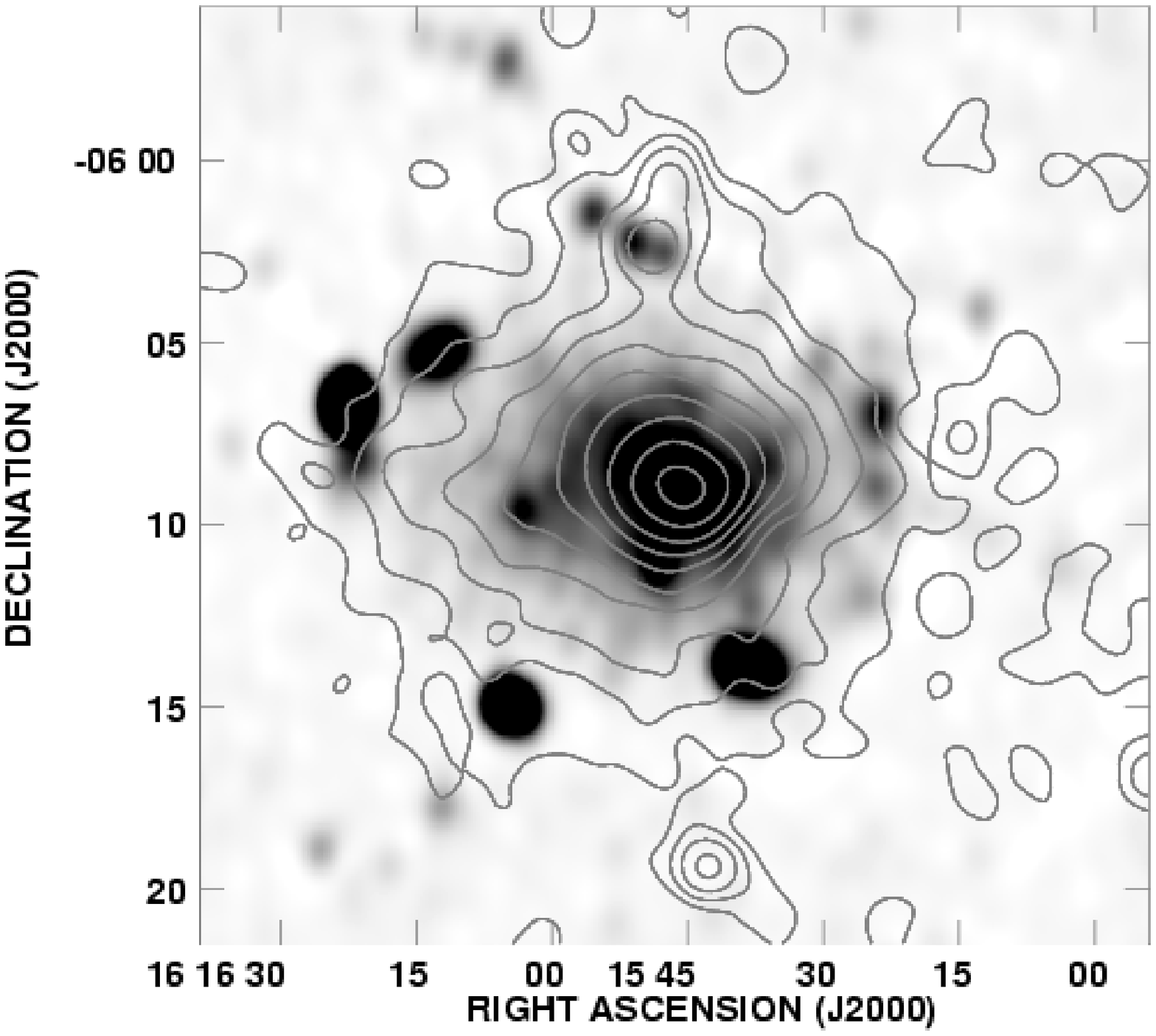}
\includegraphics[width=6cm]{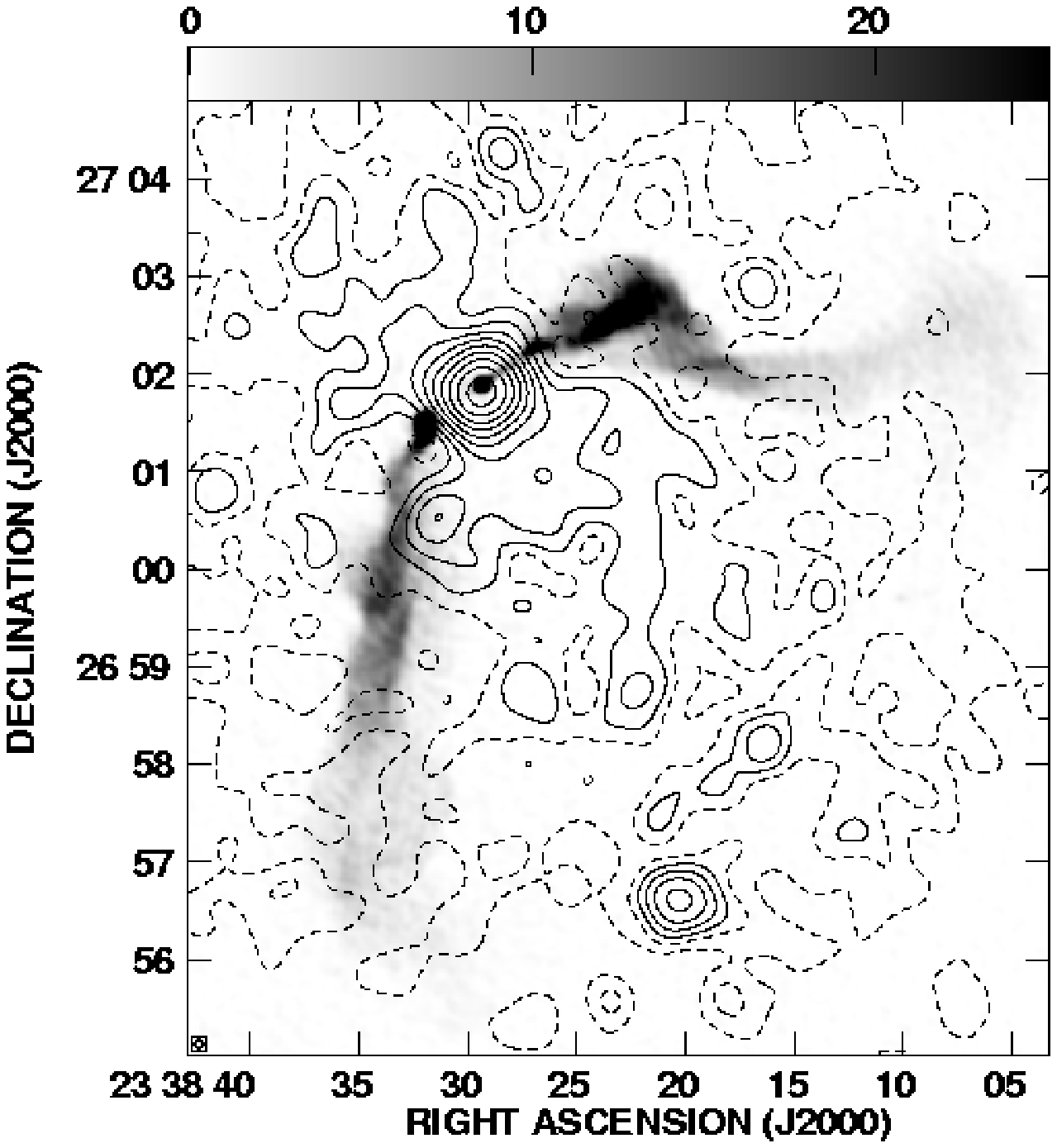}
\caption{Left: Overlay of the radio map (grey scale) onto the X-ray image (contours) for the cluster A2163 (Feretti {\it{et al.}} 2001). Right: Radio tails associated with 3C 465 in cluster A2634 (Schindler \& Prieto 1997).}
\label{fig:radio}
\end{figure}

\begin{figure}
\centering
\includegraphics[width=6cm]{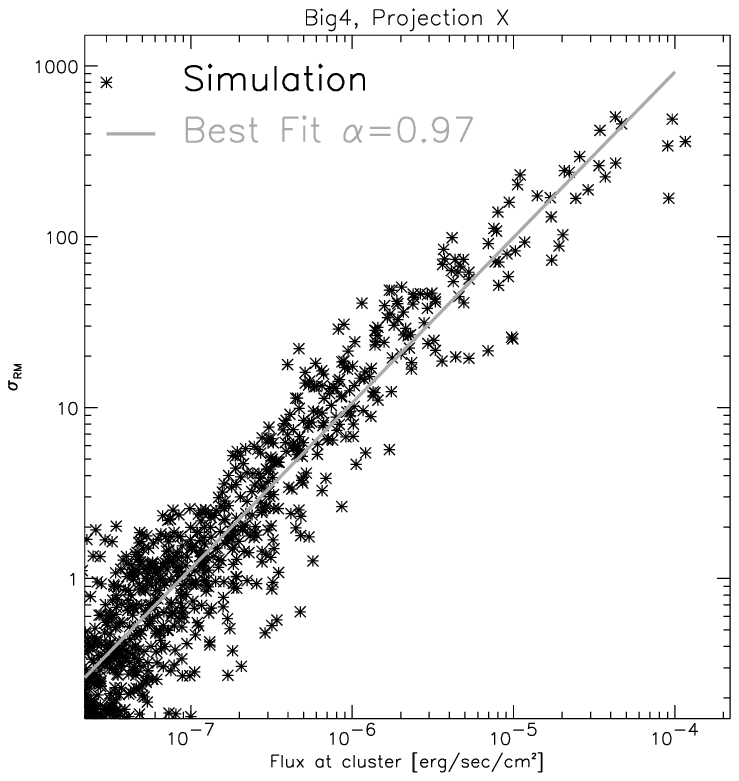}
\includegraphics[width=6cm]{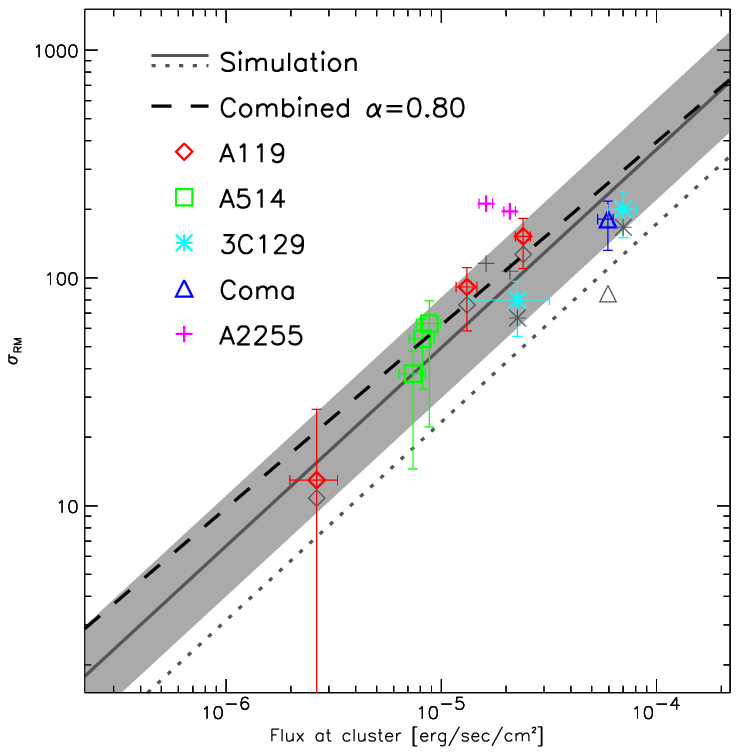}
\caption{Left: The symbols show a point by point comparison of the X-ray surface brightness and the {\it{rms}} of the synthetic rotation measure calculated from one projection of one simulated cluster taken from the ''medium'' models. Overlaid is the best fit power-law. Right: It is shown the measured correlation between the X-ray surface brightness and the {\it{rms}} of the rotation measurement. The symbols are the measurements from the clusters as indicated in the plot (from Dolag {\it{et al.}} 2001).}
\label{fig:simul}
\end{figure}

Another interesting component is relativistic particles. Their presence can be inferred from radio observation of their synchrotron emission. Radio emission has been found in many galaxies clusters. Two different kind of radio emission can be found: diffuse emission and emission associated with galaxies. In several clusters diffuse radio emission could be detected ({\it{e.g}} Giovannini {\it{et al.}} 1999~\cite{Gio}). If this emission is located in the central parts and has roughly spherical shape, it is called radio halo (see fig.\ref{fig:radio} Feretti {\it{et al.}} 2001~\cite{Fer01}). In other clusters the radio emission is situated in the outer parts and has usually elongated shapes. These sources are call relics. The origin of the radio halos is still being debated and recent suggestions point out that cluster mergers may play a crucial role for the formation and energetics of these sources.
The radio emission associated with galaxies (see fig.\ref{fig:radio} Schinlder \& Prieto 1997~\cite{SchPr}) is polarized. A magnetized plasma as the ICM it is, has the property to rotate the plane of polarization of the radiation an angle $\phi$ proportional to the observed wavelength ($\phi=RM\lambda^{2}$). Observations of the rotation measure ($RM$) of sources in or behind a cluster provide the possibility to determine the cluster magnetic field. Typically values between 0.1$\mu$G up to few $\mu$G are found (Feretti {\it{et al.}} 1995~\cite{Fer95}).
Related with the study of magnetic fields in clusters of galaxies Dolag {\it{et al.}} (2001)~\cite{Dol} have found an interesting correlation between the magnetic field and the intracluster gas density both, in their MHD simulations and in observational data taken from literature (see fig.\ref{fig:simul}). The results rule out a magnetic field that is constant within a cluster, decreasing with radius almost in the same way as the ICM density decreases with radius.

\section{Distribution of Dark and Baryonic Matter}
The mass fraction of the intra-cluster gas is not negligible. It supposes about 10-30\% of the total cluster mass. The mass in the galaxies is much smaller, about 3-5\%. The major fraction of the mass in the cluster is not visible and therefore called dark matter. In this way, measurements of the total mass of the clusters indirectly provide information on the amount and the distribution of the dark matter (see the mass profile of Abell 2199 in fig.\ref{fig:mass}).

\begin{figure}
\centering
\includegraphics[width=7cm]{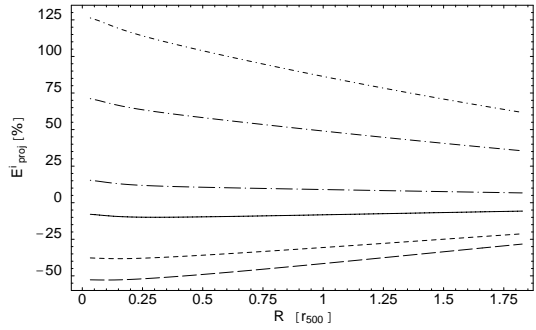}
\caption{The relative errors $E^{i}_{proj}(R)$ for the projected mass estimates for A2390 plotted for the 6 triaxial models: very compressed along the line of sight, moderately compressed and prolate shapes (dashed-dotted lines, with increasing segment length), oblate (solid line) and moderately and very elongated shapes (dashed lines with increasing segment length) (from Piffaretti{\it{et al.}} in prep.)}.
\label{fig:ellip}
\end{figure}

There are three independent methods to estimate the total cluster mass:
\begin{itemize}
\item 
The X-ray emitting gas can be used to trace the cluster potential. With the assumption of hydrostatic equilibrium and spherical symmetry cluster masses can be derived directly from X-ray observations through the gas density gradient, the gas temperature gradient and the gas temperature itself:
\begin{equation}
M(<r)=-\frac{kr}{\mu
m_{p}G}T_{gas}(r)\left (\frac{d\ln \rho_{gas}(r)}{d\ln r}+\frac{d\ln T_{gas}(r)}{d\ln r}\right).
\label{eq:mass}
\end{equation}

The improved spatial and spectral resolution of the new X-ray satellites (CHANDRA and XMM) provide very accurate measurements of cluster mass.

\begin{figure}
\centering
\includegraphics[width=7cm]{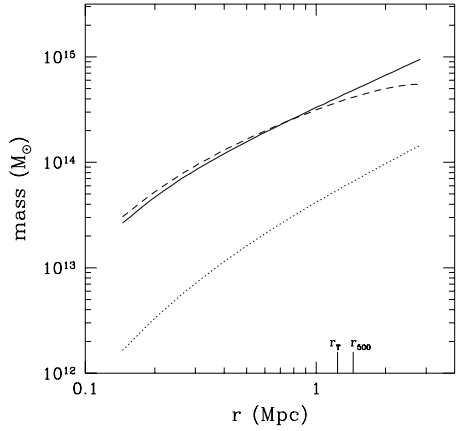}
\caption{Integrated mass profile of cluster A2199 using the X-ray method. Shown are the gas mass (dotted line), the total gravitating mass assuming isothermal cluster gas (solid line) and assuming a gradient of temperature (dashed line) (from Castillo-Morales \& Schindler in prep.).}
\label{fig:mass}
\end{figure}

\item
A second method is based on the gravitational lensing effect: light from galaxies behind clusters is deflected by the large mass of the cluster, so that we see distorted images of these galaxies. The giant arcs (strong lensing) as well as the statistical distortion of all background objects (weak lensing) can be used for the mass determination (Wambsganss 1998~\cite{Wa}; Hattori {\it{et al.}} 1999~\cite{Hat}; Mellier 1999~\cite{Me}).
\item
And the third and oldest method that uses the spectroscopic observations and determines the mass from the velocity distribution of cluster galaxies with the assumption of virial equilibrium.
\end{itemize}
For many clusters these three methods have been applied to calculate the total cluster mass. In some clusters there is agreement between the mass derived with the different methods, but in other clusters there are discrepancies typically with the following relation:
\begin{equation}
M_{X-ray} \approx M_{velocity\hspace{0.1cm}dispersion} \lesssim M_{weak\hspace{0.1cm}lensing} \lesssim M_{strong\hspace{0.1cm}lensing}
\end{equation}
with differences sometimes as high as a factor 2-3 between X-ray and strong lensing mass estimations. Between the possible explanations suggested to understand these discrepancies (non-equilibrium configurations, projection effects, multi-phase medium...), the assumption of spherical symmetry has been studied. Piffaretti {\it{et al.}} in prep. have compared the X-ray masses using the spherical model with the projected X-ray masses assuming different ellipticities models along the line of sight (see fig.\ref{fig:ellip} where the relative errors $E^{i}_{proj}(R)=\frac{M^{sph}_{proj}(R)-M^{i}_{proj}(R)}{M^{i}_{proj}(R)}$ are represented in \%). Negative relative errors imply underestimations of the total mass if spherical symmetry is assummed: this is the case if the cluster is elongated. Overestimations are found for compressed clusters. Therefore, the elongation along the line of sight could resolve the discrepancy between X-ray and lensing mass even if the difference between the two masses is as high as a factor of 2. 

\begin{figure}
\centering
\includegraphics[width=6cm]{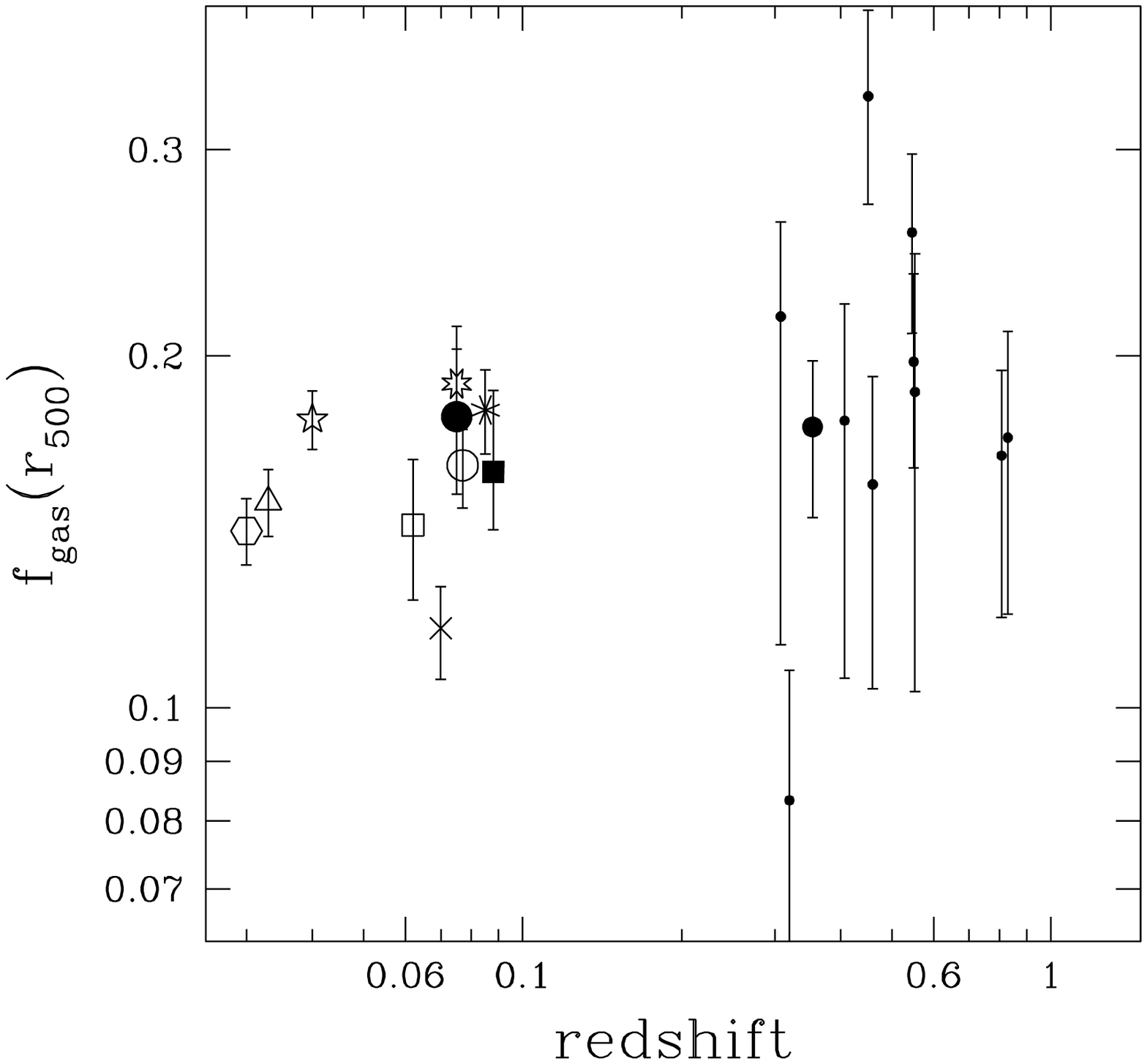}
\includegraphics[width=6cm]{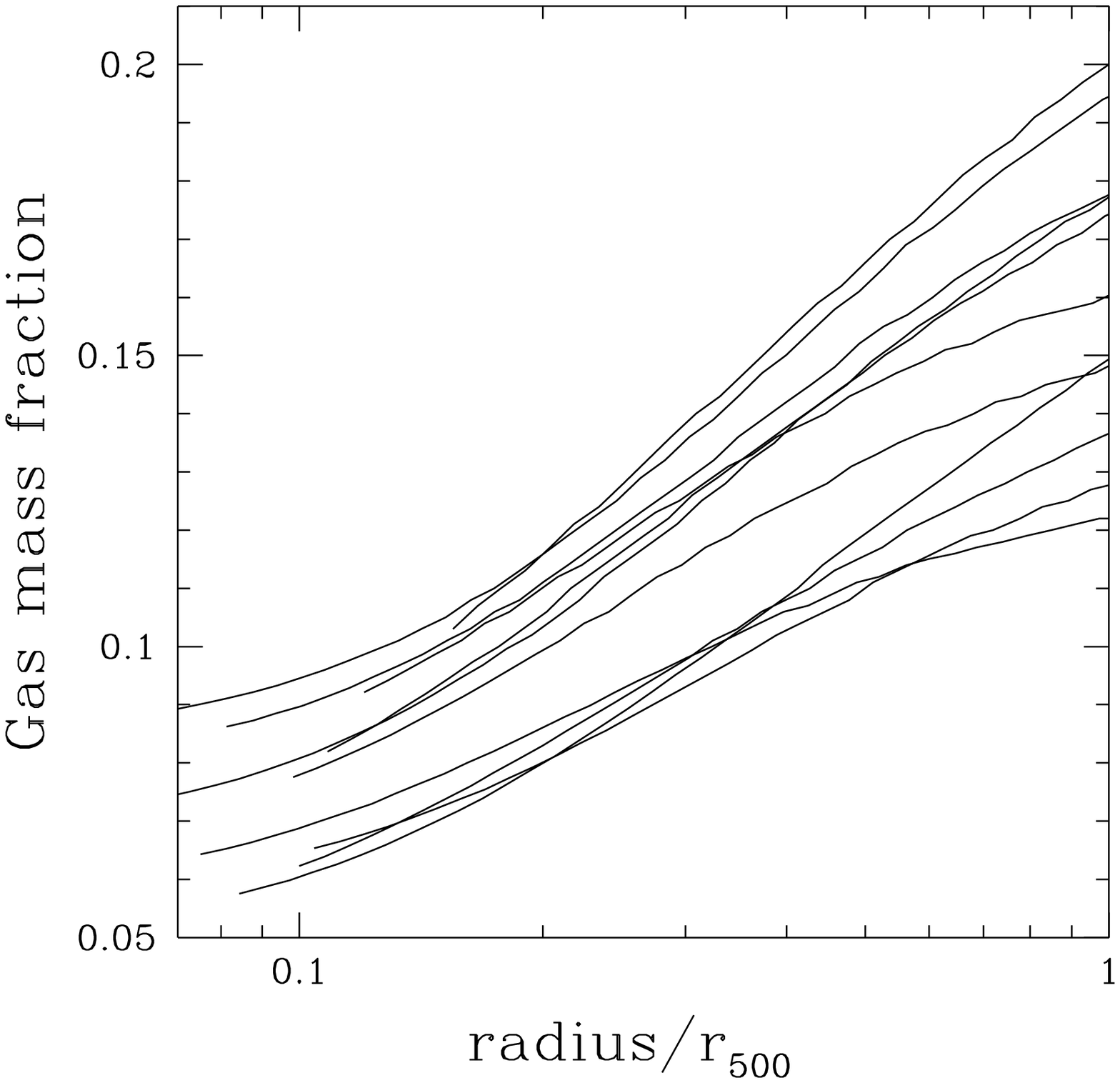}
\caption{Gas mass fraction in galaxy clusters versus redshift (left panel). The point at z=0.35 comes from CHANDRA observation, all the others use a combination of ROSAT and ASCA data. In the right panel is shown the gas mass fraction profiles derived for the nearby cluster sample (from Castillo-Morales \& Schindler, in prep.).}
\label{fig:fgasradz}
\end{figure}

The ratio of visible to total mass is a measure for the baryon fraction $\Omega_{baryon}$. Current measurements are in contradiction with primordial nucleosynthesis for $\Omega=1$ (White {\it{et al.}} 1993~\cite{Wh}), which is therefore an important hint for a low $\Omega$ universe. Several groups determined gas mass fractions from X-ray observations in samples of nearby and distant clusters, {\it{e.g}} Mohr {\it{et al.}} (1999)~\cite{Mo}: $f_{gas}=0.21$, Ettori \& Fabian (1999)~\cite{EtFab}: $f_{gas}=0.17$, Arnaud \& Evrard (1999)~\cite{ArEv}: $f_{gas}=0.16-0.20$, Schindler (1999)~\cite{Sch99}: $f_{gas}=0.18$, Castillo-Morales \& Schindler (in prep.): $f_{gas}=0.14$ (see fig.\ref{fig:fgasradz}). All these values depend on the radius where the mass fraction is determined, because the gas mass fraction increases slightly with radius (see fig.\ref{fig:fgasradz}). Therefore the gas mass fractions need to be calculated for equivalent radii. In the above mentioned analyses the mass was determined within a radius $r_{500}$ from the cluster center. This radius encompasses a volume that has a density of 500 $\times$ the critical density of the universe $\rho_{crit}$.

To determine $\Omega_{m}$, the gas mass fraction $f_{gas}$ must be compared to the baryon fraction density in the universe $\Omega_{B}$. Burles \& Tytle (1998a,b)~\cite{Bura}\cite{Burb} found $\Omega_{B}\lesssim 0.08$ from primordial nucleosynthesis. The ratio of the baryon density and the gas mass fraction yields and upper limit for the matter density $\Omega_{m}< \frac{\Omega_{B}}{f_{gas}}\approx 0.3-0.4$. In these analyses only the mass in the intra-cluster gas was taken into account. Baryons in the galaxies were neglected. If they were to be included, the baryon fraction would increase slightly and hence ever more strigent constraints on $\Omega_{m}$ could be placed.

\begin{figure}
\centering
\includegraphics[width=7cm]{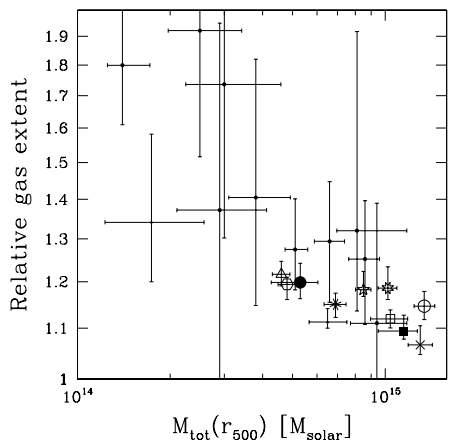}
\caption{Gas extent relative to the dark matter extent expressed as the ratio of gas mass fractions at large radius and small radius (small points represent the distant cluster sample from Schindler 1999 and the larger symbols, the nearby cluster sample by Castillo-Morales \& Schindler in prep.). The gas is relatively more extended in less massive clusters, which is an indication of additional, non-gravitational heating processes.}
\label{fig:E}
\end{figure}

The relative distribution of gas and dark matter within a cluster gives also useful information on physical processes occurring in clusters. The fact that the gas is more extended than the dark matter - an effect which is stronger for the less massive cluster (see fig. \ref{fig:E}) - can be explained by additional (non-gravitational) heating processes, which are more efficient in less massive clusters (Schindler 1999; Ponman {\it{et al.}} 1999). This gives in principle hints on when and how the heat is produced and hence information about cluster formation.
For cosmology and cluster formation it is very interesting to know how the baryon fraction in clusters changes with time. In fig.\ref{fig:fgasradz} the gas mass fraction is plotted versus the redshift. It is clear that with the previous ROSAT and ASCA observations it was hard to see any evolution due to the large observational uncertainties. With XMM and CHANDRA we will be able to measure gas mass fractions out to redshifts of almost unity with good accuracy.

In fig. \ref{fig:fgasradz} we see that even with the old observations the gas mass fractions varies considerably from cluster to cluster. These different baryon fraction reflect probably the very early distribution of baryonic and non-baryonic matter and they are therefore of high interest for cosmology.
\begin{figure}
\centering
\includegraphics[width=6cm]{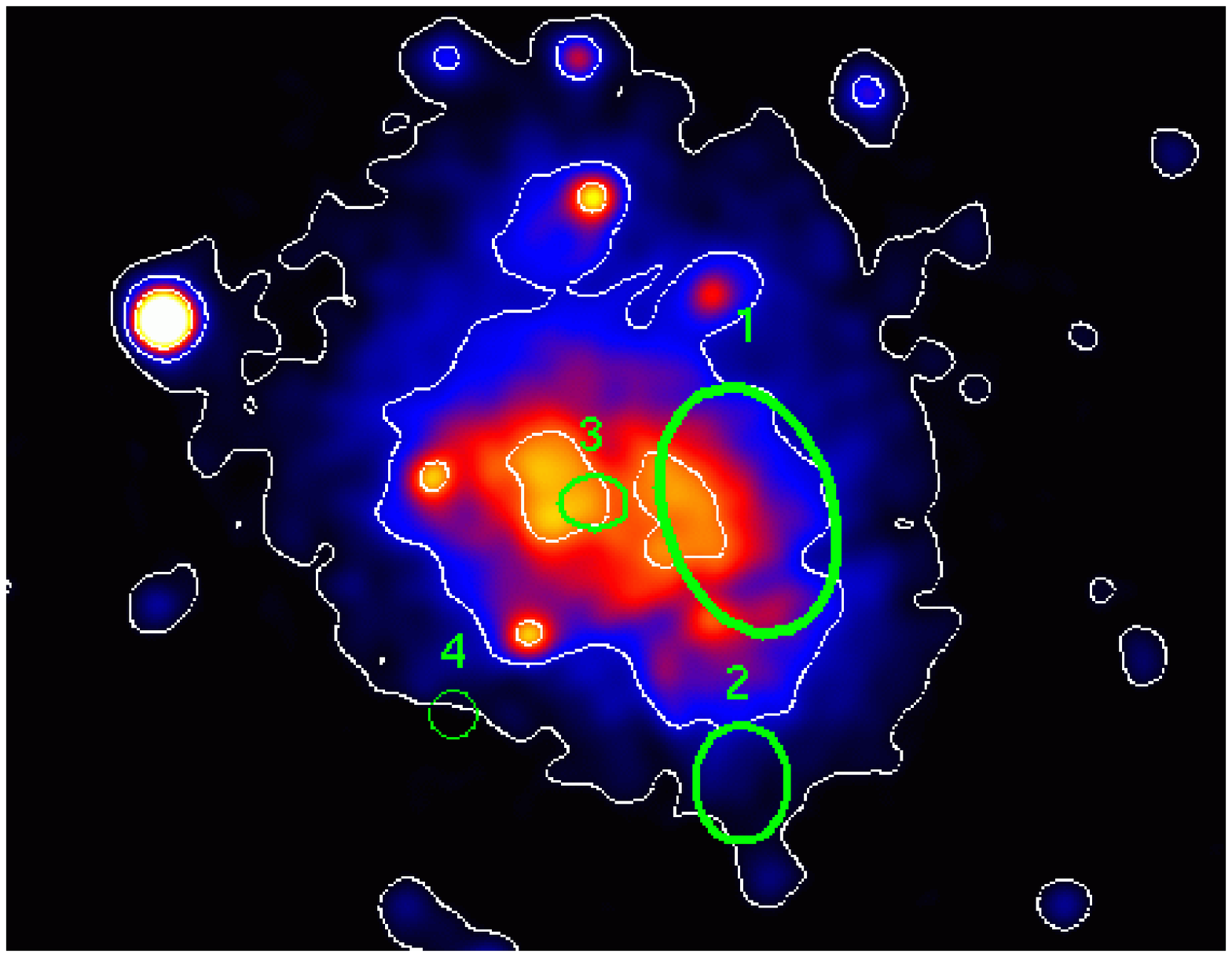}
\includegraphics[width=6cm]{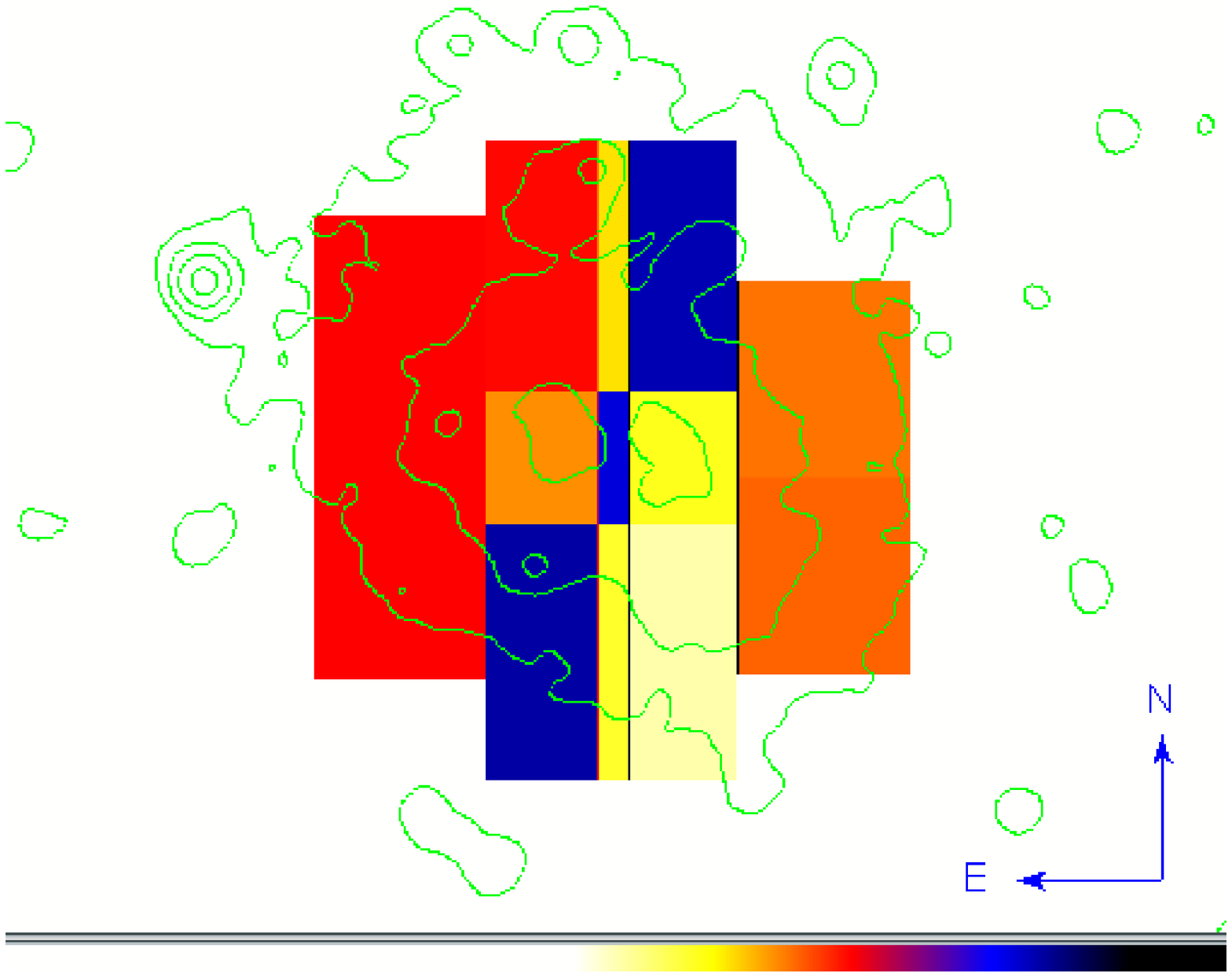}
\caption{Left: Adaptively smoothed image of the galaxy cluster CL~0939+4713. The image is a mosaic of the three MOS1, MOS2 \& pn cameras. The core of CL~0939+4713 does not show a simple single central structure, but is composed by two subclumps. The green circles show the regions of density of all galaxies that until today have been proved to belong to the cluster. Right: Cluster temperature map; superimposed are linearly spaced X-ray contours. A trend of higher temperatures is observed in the central region, located between the two peaks in the X-ray emission. The high temperature region also extends in the NW and SE directions (from De Filippis {\it{et al.}} A\&A submitted)}
\label{fig:cl0939}
\end{figure}
\section{Distant Clusters of Galaxies}

Clusters of galaxies can be used to study how structures form on large scales. The formation and evolution of clusters depends very sensitively on cosmological parameters like the mean matter density in the universe $\Omega_{m}$ (Thomas {\it{et al.}} 1998~\cite{Tho}; Jenkins {\it{et al.}} 1998~\cite{Jen}; Beisbart {\it{et al.}} 2001~\cite{Be}). Therefore it is of great importance to determine the dynamical state of clusters at different redshifts and hence the study of high redshift systems ($z>2$). But the detection of (proto)clusters at these redshifts using conventional optical and X-ray techniques is not easy. In a recent imaging and spectroscopy study with the Very Large Telescope, Venemans {\it{et al.}} (2002)~\cite{Ve} have found a structure of 20 Ly$\alpha$-emitting galaxies around the high redshift radio galaxy TN J1338-1942. The overdensity of this protocluster is on the order of 15 compared to field samples. Their results demonstrate that by z=4.1, megaparsec-scale structure had already formed.

The high sensitivity of XMM makes it the ideal instrument to find and study distant galaxy clusters in order to investigate evolutionary effects. In fig.\ref{fig:cl0939} is shown an XMM observation of the intermediate distant cluster CL 0939+4713 ($z=0.41$)(De Filippis {\it{et al.}} A\&A submitted). The X-ray image shows pronounced substructures. There are two main subclusters which have even some internal structure. This is an indication that the cluster is a dynamically young system. This conclusion is supported by the temperature distribution (see fig.\ref{fig:cl0939}): a hot region is found between the two main subclusters indicating that the cluster is in the process of a major merger, in which the two subclusters will probably collide in a few hundreds Myr. The intra-cluster gas of CL 0939-4713 shows variations of the metal abundances having the optically richer subcluster a somewhat higher metallicity. The detection of such a metallicity variations gives hints on the metal enrichment processes.

\section{Summary}

Clusters of galaxies can be used for very different types of cosmological tests. The X-ray astronomy is entering in a new era thanks to the new capabilities of the X-ray telescopes, {\it{e.g}} XMM and Chandra. The high spatial resolution of Chandra (1'') allows to investigate cluster morphology and derive very accurate profiles. Distant clusters can now be investigate thanks to the high sensitivity of XMM. These two satellites are capable of performing spatially resolved spectroscopy with high accuracy, which is necessary for temperatures maps, mass derivations and metallicity distributions. The combination of these data with observations in other wavelength will open new horizons in cosmology.

%

\begin{thebibliography}{}

\bibitem {Dre} A. Dressler, ApJ {\bf{236}} (1980) 351
\bibitem {Fuk} Y. Fukazawa, K. Makishima, T. Tamura, {\it{et al.}}, PASJ {\bf{50}} (1998) 187
\bibitem {Gio} G. Giovannini, M. Tordi, L. Feretti, New Astronomy {\bf{4}} (1999) 141
\bibitem {Fer01} L. Feretti, R. Fusco-Femiano, G.  Giovannini, F. Govoni, A\&A {\bf{373}} (2001) 106
\bibitem {SchPr} S. Schindler \& M.A. Prieto, A\&A {\bf{327}} (1997) 37
\bibitem {Fer95} L. Feretti, D. Dallacasa, G. Giovannini, A. Tagliani, A\&A {\bf{302}} (1995) 680
\bibitem {Dol} K. Dolag, S. Schindler, F. Govoni, L. Feretti, A\&A {\bf{378}} (2001) 777
\bibitem {Wa} J. Wambsganss, Living Reviews in Relativity {\bf{1}} (1998) 12
\bibitem {Hat} M. Hattori, J.P. Kneib, N. Makino, Prog. Theor. Phys., Supplement {\bf{133}} (1999) 1
\bibitem {Me} Y. Mellier, ARA\&A {\bf{37}} (1999) 127
\bibitem {Wh} S.D.M. White, J.F Navarro, A.E. Evrard, C.S. Frenk, Nature {\bf{366}} (1993) 429
\bibitem {Mo} J.J. Mohr, B. Mathiesen, A.E. Evrard, ApJ {\bf{517}} (1999) 627 
\bibitem {EtFab} S. Ettori \& A.C. Fabian, MNRAS {\bf{305}} (1999) 834 
\bibitem {ArEv} M. Arnaud \& A.E Evrard, MNRAS {\bf{305}} (1999) 631
\bibitem {Sch99} S. Schindler, A\&A {\bf{349}} (1999) 435
\bibitem {Bura} S. Burles \& D. Tytle, ApJ {\bf{499}} (1998a) 699
\bibitem {Burb} S. Burles \& D. Tytle, ApJ {\bf{507}} (1998b) 732
\bibitem {Pon} T.J. Ponman, D.B. Cannon, J.F. Navarro, Nature {\bf{397}} (1999) 135
\bibitem {Tho} P.A. Thomas, J.M. Colberg, H.M.P. Couchman, {\it{et al.}}, MNRAS {\bf{296}} (1998) 1061 
\bibitem {Jen} A. Jenkins, C.S. Frenk, F.R. Pearce, P.A. Thomas, {\it{et al.}}, ApJ {\bf{499}} (1998) 20
\bibitem {Be} C. Beisbart, R. Valdarnini, T. Buchert, A\&A {\bf{379}} (2001) 412 
\bibitem {Ve} B.P. Venemans, J.D. Kurk, G.D. Miley, {\it{et al.}}, ApJ {\bf{569}} (2002) 11 

\end{thebibliography}
\end{document}